\documentclass[aps,prl,reprint,superscriptaddress]{revtex4-1}

\usepackage{graphicx}
\usepackage{rotating}
\usepackage{amssymb}    
\usepackage{amsmath}   
\usepackage{epsfig}
\usepackage[normalem]{ulem}   
\usepackage{bm}   
\usepackage{color}

\date{\today}

\begin{document}

\title{Asymmetry-induced order in multilayer networks}

\author{Everton S. Medeiros}
\email{medeiros@tu-berlin.de}
\affiliation{Institut f\"ur Theoretische Physik, Technische Universit\"at Berlin, Hardenbergstra\ss e 36, 10623 Berlin, Germany}
\author{Ulrike Feudel}
\affiliation{Institute for Chemistry and Biology of the Marine Environment, Carl von Ossietzky University Oldenburg, 26111 Oldenburg, Germany}
\author{Anna Zakharova}
\email{anna.zakharova@tu-berlin.de}
\affiliation{Institut f\"ur Theoretische Physik, Technische Universit\"at Berlin, Hardenbergstra\ss e 36, 10623 Berlin, Germany}

\begin{abstract}
Symmetries naturally occur in real-world networks and can significantly influence the observed dynamics. For instance, many synchronization patterns result from the underlying network symmetries, and high symmetries are known to increase the stability of synchronization. Yet, here we find that general macroscopic features of network solutions such as regularity can be induced by breaking their symmetry of interactions. We demonstrate this effect in an ecological multilayer network where the topological asymmetries occur naturally. These asymmetries rescue the system from chaotic oscillations by establishing stable periodic orbits and equilibria. We call this phenomenon \textit{asymmetry-induced order} and uncover its mechanism by analyzing both analytically and numerically the absence of dynamics on the system's synchronization manifold. Moreover, the bifurcation scenario describing the route from chaos to order is also disclosed. We demonstrate that this result also holds for generic node dynamics by analyzing coupled paradigmatic R\"ossler and Lorenz systems.
\end{abstract}
\maketitle

\section{Introduction}

Symmetries play an essential role in the dynamics of complex systems. For instance, in complex networks, highly symmetric connections are known to support synchronization \cite{Aguiar2011,Nicosia2013} and the underlying network symmetries define stable synchronization patterns \cite{Pecora2014, Poel2015, Krishnagopal2017,Zakharova2020,Berner2020}. However, many living and engineered dynamical systems are inherently asymmetric, i.e., their topological symmetry is broken. In soft matter physics \cite{Loos2019, Loos2020} and quantum optics \cite{Metelmann2015,Fang2017} such interactions are referred to as non-reciprocal. Recently reported examples are provided by colloidal particles \cite{Lavergne2019}, active matter models describing self-propelled motion \cite{Speck2019}, novel quantum devices, such as directional amplifiers \cite{Shen2018}. In \textit{multilayer networks}, offering a realistic representation of real-world systems \cite{Boccaletti2014, Kivela2014,Domenico2016,Pilosof2017,Semenova2018, Danziger2019,Hutchinson2019,Schuelen2020}, asymmetric interactions also emerge naturally. The examples range from financial networks \cite{Brummitt2015} to disease spreading processes \cite{Wang2014} and gene regulation \cite{Klosik2017}. One of the key problems concerning dynamical regimes of the real-world systems and networks is the conditions required to prevent instabilities such as unpredictability and amplitude variability of their solutions \cite{Ott1990,Kocarev1996,Scholl2008,Glendinning1994,HegedHus2020}.

In spite of the growing interest in various fields of research in the topic of multilayer networks and topological asymmetries, little is known about the interplay of these two concepts and their impact on the system's dynamical regimes. While synchronization and its stability in systems with asymmetries have been previously considered \cite{Nishikawa2016,Hart2019,Molnar2020,Sugitani2021,Molnar2021}, the fundamental question related to general features of emerging solutions and its interconnection with the structural asymmetries remains to be understood.

In this work, we fill this gap by establishing a relation between naturally occurring topological asymmetries in multilayer networks and the state-space characteristics of the corresponding network solutions. We show that to induce stable time-periodic or equilibria solutions one has to break the symmetry of the interaction between the network layers. We call this counter-intuitive effect \textit{asymmetry-induced order} (AiO) and demonstrate this phenomenon exploiting an ecological network where the topological asymmetry occurs naturally due to the dispersal abilities of species in different landscapes. To analyze the solution's features, we calculate Lyapunov spectra and identify bifurcations mediating the observed transitions. Moreover, we investigate the sensitivity of AiO to the level of asymmetry in the network characterized by the number of asymmetric inter-layer links and to the amount of heterogeneity of the individual network elements in the layers. Next, we explain the onset of AiO as a result of mutual inter-layer disturbances preventing the system's complete synchronization. The absence of dynamics on the synchronization manifold caused by layer asymmetries is demonstrated analytically and numerically by computing the system synchronization error. Finally, we show that AiO occurs in a general class of networks of diffusively coupled elements by analyzing paradigmatic models such as R\"ossler and Lorenz systems.

\section{Networked Ecological Model}

Here, we choose an ecological system for our detailed analysis motivated by the natural way that the multilayer structure \cite{Pilosof2017,Hutchinson2019} and interaction asymmetries appear in these systems. Specifically, the species traits for developing efficient movement strategies are essentially the outcome of adaptative evolutionary processes resulting in biomechanical, physiological, and behavioral advantages \cite{Shepard2013}. However, the species movement performance may also vary according to the characteristics of the landscape that is being traversed \cite{Shepard2013}. Landscape factors such as incline, substrate type, vegetation, current speed, or direction also play a fundamental role in the overall species' dispersal capability \cite{Wilson2012,Shepard2013}. Hence, within its foraging distances, an animal species may go across landscapes divided into layers distinguished by such conditions \cite{Dunford2020}. Consequently, the dispersal ability is constantly changing following the landscape conditions. Given the crucial role of dispersal for the species' population dynamics \cite{Tromeur2016,Gramlich2016}, the occurrence of landscapes with different dispersal properties has deep consequences. Hence, we identify the landscapes as layers in a multilayer network and assign different dispersal properties to each layer.

We consider a food-web model to specify the population dynamics on ecological patches containing species of three different trophic levels: vegetation ($u$), prey species ($v$), and predators ($w$). The spatial organization of the model is provided by connecting $N$ heterogeneous patches forming an ecological landscape which, in turn, is partitioned in two layers corresponding to landscapes offering different dispersal features. Each of such one-dimensional layers is defined by a distinct migration capability of the animal species. The system of equations describing the dynamics of the patch $i$ with $i = 1, \dots, N_{\mu}$ in the layer $\mu$ with $\mu=\{1,2\}$ is given by:
\begin{eqnarray}
\label{FW}
\nonumber
\dot{u}_{i\mu} &=& u_{i\mu}(1-u_{i\mu})-f_1(u_{i\mu})v_{i\mu},\\ [5pt]
\dot{v}_{i\mu} &=& v_{i\mu}\left[f_1(u_{i\mu})-d_v\right] -f_2(v_{i\mu})w_{i\mu} + \Delta_{i\mu \mu'}v_{i\mu},\\ [5pt]
\dot{w}_{i\mu} &=& w_{i\mu}\left[f_2(v_{i\mu})-d_{wi\mu} \right] + \Delta_{i\mu \mu'}w_{i\mu}.
\nonumber
\end{eqnarray}
The vector ${\bf r}_{i\mu}(t)=(u_{i\mu}(t), v_{i\mu}(t), w_{i\mu}(t))$ defines the state-space of each ecological patch. The population dynamics of isolated patches was first introduced in Ref. \cite{Hastings1991}. The transfer of biomass across the trophic levels is controlled by a Holling type-II response specified by $f_k(x)=\alpha_kx/(1+\beta_kx)$ with $k=\{1,2\}$, depending on the species trophic level. The constant $\alpha_k$ parameterizes the saturation level of the trophic interaction, while $\beta^{-1}_k$ specifies the half-saturation constant of the consumption rate per unit of the resource ($k=1$ for vegetation and $k=2$ for prey). We keep $\alpha_k$ and $\beta_k$ fixed in our study at the values $\alpha_1=5.0$, $\beta_1=3.0$, $\alpha_2=0.1$, and $\beta_2=2.0$ \cite{Hastings1991}. The death rates for herbivores and predators are controlled by the parameters $d_v=0.35$ and $d_{wi\mu}$, respectively. The values of $d_{wi\mu}$ can vary among the patches and layers providing a source of heterogeneity in the system. This parameter mismatch is defined as $d_{wi\mu}= 0.00650 + \gamma \times (i+(\mu-1)N_{\mu})$ with $i=1,\dots,N_{\mu}$ and $\mu=1,2$. The mismatch intensity $\gamma$ is fixed at $\gamma=0.00015$ throughout the study. 

\begin{figure}[!htp] 
\centering
\includegraphics[width=7.1cm,height=3.5cm]{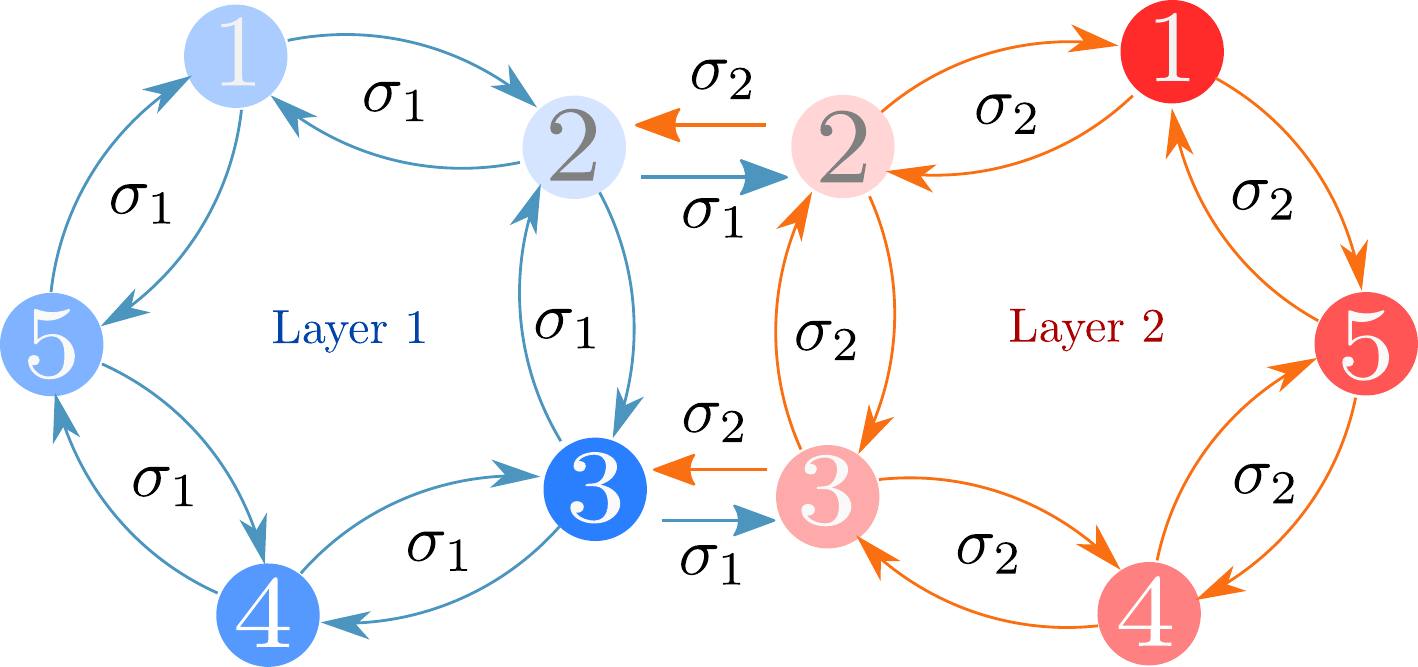}
\caption{Schematic representation of a two-layer network with two asymmetric inter-layer links.}
\label{figure_1}
\end{figure}

The connections among patches are controlled by the term $\Delta_{i\mu \mu'}x_{i\mu}=\delta_{i\mu}x_{i\mu}+\delta_{i\mu \mu'}x_{i\mu}$ in Eq. (\ref{FW}) where $\delta_{i\mu}x_{i\mu}$ and $\delta_{i\mu \mu'}x_{i\mu}$ are the intra-layer and inter-layer species mobilities, respectively. For a review on network coupling functions, see ref. \cite{Stankovski2017}. For species moving inside one of the layers, the interactions are symmetric and the dispersal mechanism is modeled by $\delta_{i\mu}x_{i\mu}=\sigma_{\mu}\left[x_{(i+1)\mu}+x_{(i-1)\mu}-2x_{i\mu}\right]$. The parameter $\sigma_{\mu}$ specifies the dispersal or coupling strength. Periodic boundary conditions are adopted for the intra-layer species movement. This ring structure is a one-dimensional version of lattice networks known to satisfactory reproduce the dispersal of animal species \cite{Blasius1999}. For species moving across the layers via the patches that connect them, the dispersal mechanism is expressed as $\delta_{i\mu \mu'}x_{i\mu}= \sigma_{\mu'}\mathcal{A}_{ii`}x_{i'\mu'} - \sigma_{\mu}\mathcal{A}_{ii`}x_{i\mu}$. The subscripts $i'$ and $i$ specify the patch indexes of opposite layers $\mu'$ and $\mu$, respectively. The inter-layer connections between $i`$ and $i$ are specified by an adjacency matrix $\mathcal{A}$ with $N_{\mu} \times N_{\mu`}$ elements $\mathcal{A}_{ii`}$. Since both layers are of the same size $N_{\mu}=N_{\mu'}$ and only one inter-layer link is allowed per patch, we attach the inter-layer links to patches with the same intra-layer index. Therefore, the elements of $\mathcal{A}$ are $\mathcal{A}_{ii`}=0$ for $i` \neq i$ and $\mathcal{A}_{ii}=1$ for the cases in which there is a connection between $i$ and $i`$, otherwise $\mathcal{A}_{ii}=0$ \cite{Mucha2010,Gomez2012}. The inter-layer connections are asymmetric, i.e., the strength of the inter-layer coupling is defined by its direction. It implies that the dispersal constant of the source layer characterizes the species movement until it reaches the corresponding patch in the neighbor layer. In Fig. \ref{figure_1}, we illustrate such topological features for layer sizes $N_1=N_2=5$ and inter-layer links at $i,i`=2,3$. The corresponding adjacency matrix is shown in the Appendix A, this matrix is adopted throughout the text. 

\begin{figure}[!htp] 
\centering
\includegraphics[width=8.5cm,height=3.6cm]{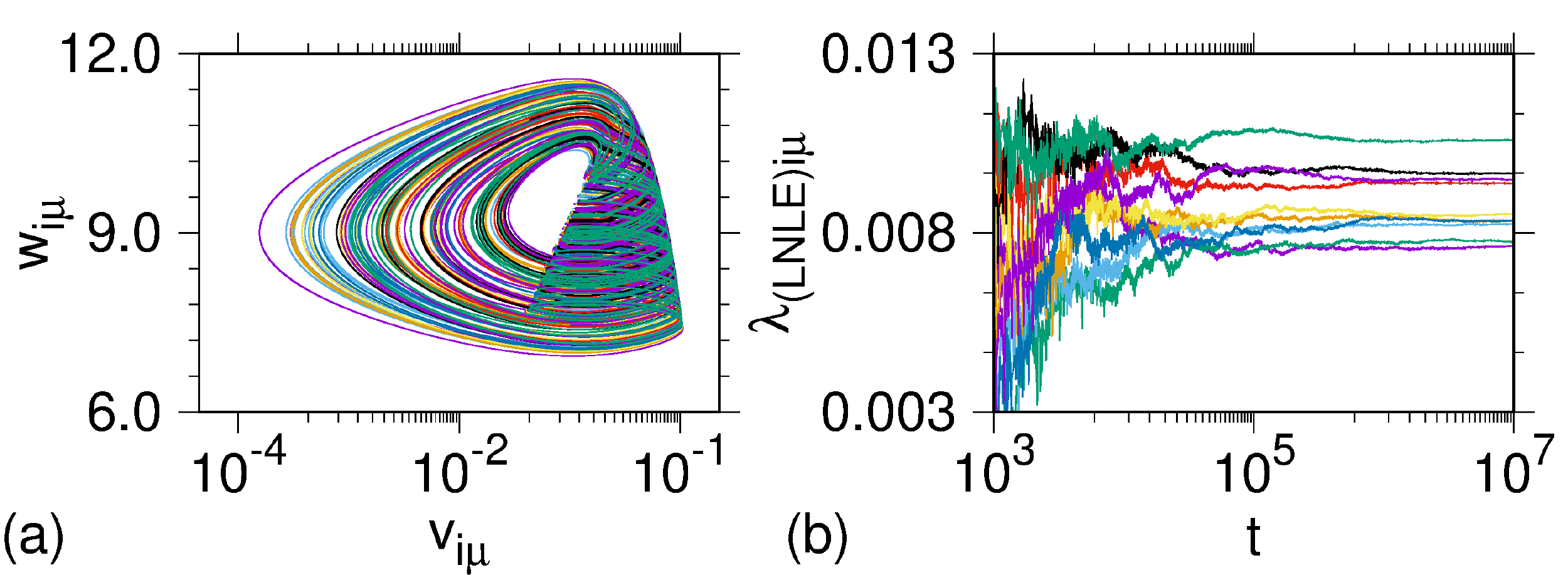}
\caption{Dynamics of uncoupled patches ($\sigma_{\mu}=0$). (a) State-space projection of the population density of species in every uncoupled patch. (b) The LNLEs ($\lambda_{(LNLE)i\mu}$) of every uncoupled patch. The colors indicate different patches.}
\label{figure_2}
\end{figure}

The solutions of the system in Eq. (\ref{FW}) are assessed by estimating their largest nonzero Lyapunov exponent (LNLE), i.e., the rate of separation of nearby trajectories in the system's state-space. A positive LNLE means that trajectories initially infinitesimally close  diverge exponentially as time passes, indicating chaotic dynamics. A negative LNLE indicates regular solutions such as limit-cycles and equilibria. We estimate the LNLE of a trajectory by following the procedure described in Ref. \cite{Wolf1985,Geist1990}. To illustrate the population dynamics occurring in the $10$ uncoupled patches ($\sigma_{\mu}=0$), we provide their state-space projection ($v_{i\mu}, w_{i\mu}$) (Fig. \ref{figure_2}(a)) and the LNLEs $\lambda_{(LNLE)i\mu}$ (Fig. \ref{figure_2}(b)). Due to heterogeneity, the patches follow different trajectories and are characterized by a different value of LNLE. However, for all patches, the LNLEs are positive indicating chaotic dynamics. An important observation in these solutions is the occurrence of large amplitude variability. 

\section{Asymmetry-Induced Order}

\begin{figure}[!htp] 
\centering
\includegraphics[width=8.5cm,height=3.7cm]{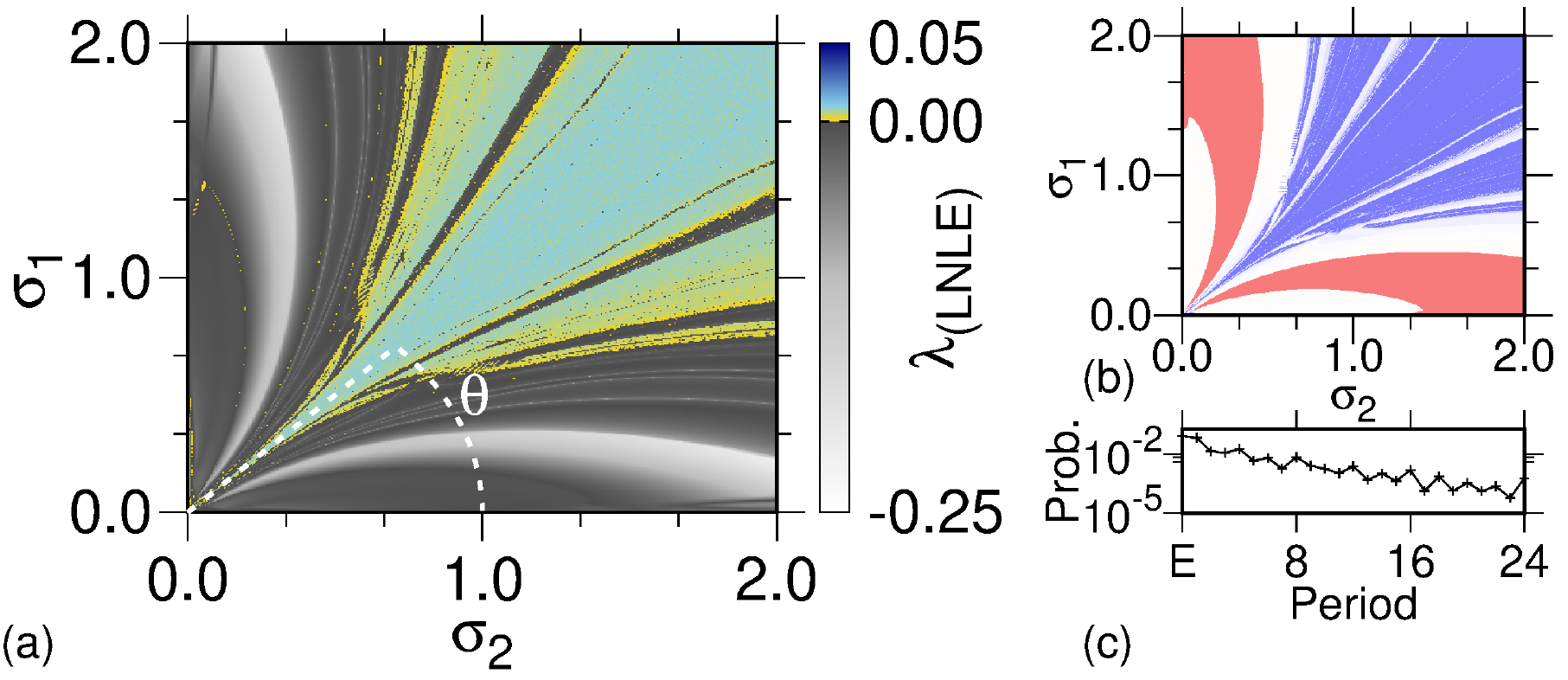}
\caption{(a) Stability diagram in the ($\sigma_2$, $\sigma_1$) parameter plane. The colors represent the values of LNLEs ($\lambda_{(LNLE)i\mu}$). (b) Map of regimes in the same parameter plane. The colors indicate different regimes: equilibria (red), limit-cycles (white), and chaos (blue). (c) The probability density of equilibria (E value in the $x$-axis) and periods in the interval $[1,24]$.}
\label{figure_3}
\end{figure}

In the coupled system, we investigate the consequences of landscapes offering two distinct conditions for preys and predators species to disperse. By computing the LNLEs of solutions for a large number of combinations of the dispersal constants $\sigma_1$ and $\sigma_2$, we look for $\lambda_{(LNLE)i\mu}<0$ indicating nonchaotic solutions. An intriguing observation becomes evident from the stability diagram displaying the LNLEs in the ($\sigma_2$, $\sigma_1$) parameter plane (Fig. \ref{figure_3}(a)). We distinguish between the dispersal combinations yielding positive values of LNLEs which correspond to chaotic solutions (yellow to blue color scale in Fig. \ref{figure_3}(a)) and the dispersal combinations leading to solutions free of internal instabilities, such as limit-cycles and equilibria (gray to white color scale in Fig. \ref{figure_3}(a)). Interestingly, we find the higher quality solutions, i.e., regular and with low amplitude variability, for asymmetric combinations of the dispersal conditions. We call this novel phenomenon \textit{asymmetry-induced order}. We refer to asymmetries as structural heterogeneities of the network in accordance to the literature on asymmetry-induced phenomena in networked systems \cite{Nishikawa2016,Hart2019,Molnar2020,Sugitani2021,Molnar2021}. In addition, we emphasize that these topological heterogeneities which we call asymmetries should not be confused with the patch heterogeneities. 

To gain more insights, we take a closer look at the solutions observed in the network. We provide in Fig. \ref{figure_3}(b) a map of regimes in the ($\sigma_2$, $\sigma_1$) parameter plane displaying the regions corresponding to limit-cycles with up to period-$24$ (white), equilibria (red), and chaotic behavior (blue). In this figure, the period of a limit-cycle in a given patch is computed from a Poincar\'e section $\Sigma$ defined as $v_{i\mu} \times w_{i\mu}$ for every local maximum of the variable $u_{i\mu}$. The period of the orbit in this section is defined as an integer $p$ such that $x_{\Sigma}(t_{\Sigma}+p)=x_{\Sigma}(t_{\Sigma})$. For the values of $\sigma_1$ and $\sigma_2$ corresponding to periodic dynamics in Fig. \ref{figure_3}(b), all patches across the network exhibit a time evolution with the same period, except for the limit of small coupling strength at which the system does not support any kind of synchronization. Moreover, we calculate the probability density of occurrence for every period as well as for equilibria (Fig. \ref{figure_3}(c)). Among the regular solutions, the highest probability of occurrence is observed for the equilibria and the limit-cycles with a lower period.

Further, we analyze the transitions between the different regimes observed in the network and the corresponding bifurcation mechanisms. First, we define the angle $\theta=\arctan({\sigma_1/\sigma_2})$ as shown in Fig. \ref{figure_3}(a). Complete symmetry between the layers is obtained for $\theta=\pi/4$, chaotic solutions are predominant regardless of the dispersal intensity in the layers. By breaking the dispersal symmetry between the layers, for instance, by decreasing $\theta$ along the dashed line in Fig. \ref{figure_3}(a), we observe the appearance of small domains of periodicity intermingled with parameter intervals corresponding to chaotic solutions. As the angle $\theta$ goes further away from $\pi/4$, we observe the occurrence of larger domains of periodicity with lower periods of oscillations until an inverse Hopf bifurcation provides an equilibrium solution. Finally, after a parameter interval corresponding to equilibria solutions, another Hopf bifurcation creates oscillations extending to the limit $\theta \approx 0$. It is important to note that the stability diagram obtained has common features of parameter spaces of systems displaying chaos \cite{Gallas1993}. 

\begin{figure}[!htp] 
\centering
\includegraphics[width=8.5cm,height=3.6cm]{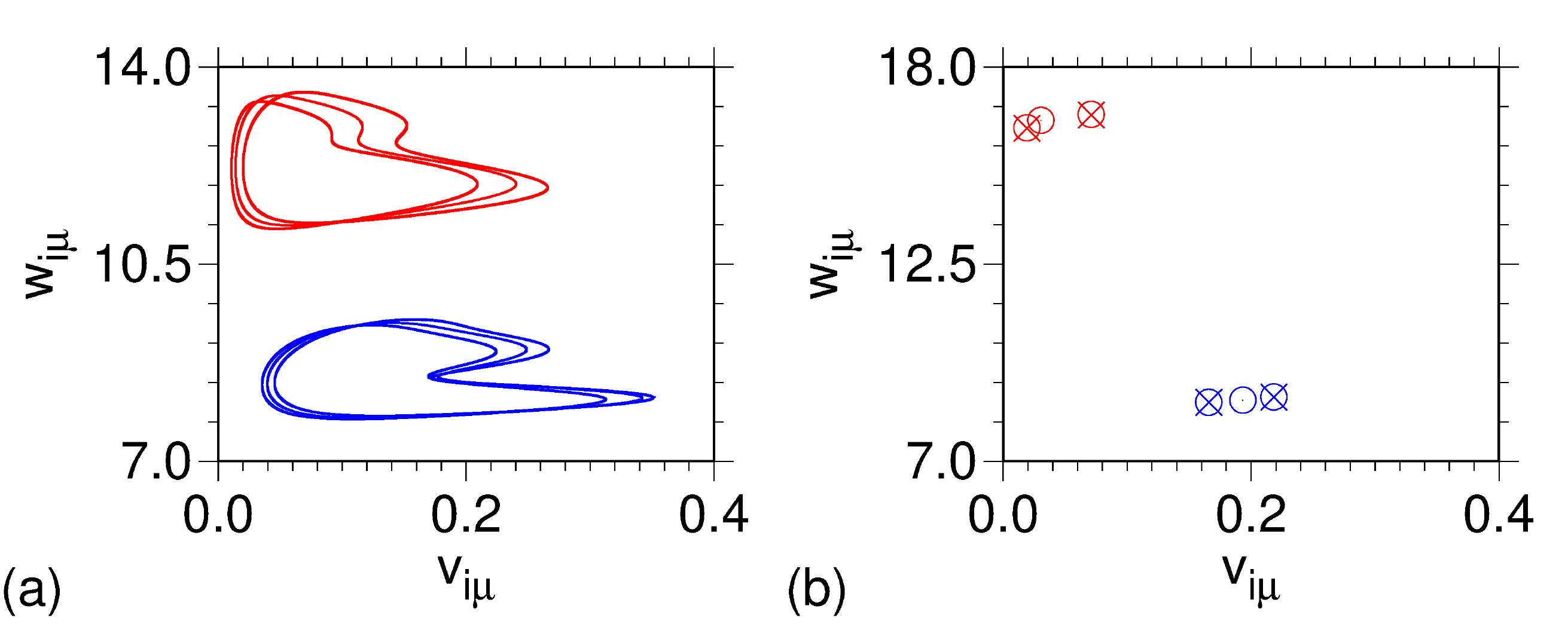}
\caption{State-space projection of the population density in the network with asymmetrically coupled layers for different level of asymmetry: (a) $\sigma_1=0.5$ and $\sigma_2=0.34$ (limit-cycles); (b) $\sigma_1=0.5$ and $\sigma_2=0.25$ (equilibria). Blue (red) color stands for patches in layer $1$ (layer $2$).}
\label{figure_4}
\end{figure}

The population density of the species moving across the landscapes with different dispersal features can be visualized in state-space projections of the system's solution. For $\sigma_1=0.5$ and $\sigma_2=0.34$, we observe that the solution is localized in two state-space domains corresponding to different layers, i.e., layer $1$ (blue curve in Fig. \ref{figure_4}(a)) and layer $2$ (red curve in Fig. \ref{figure_4}(a)). For this asymmetry level, the population density in every patch of a given layer is oscillatory following different limit-cycles. Due to the proximity of the trajectories, not all $5$ limit-cycles per layer can be discerned in this figure. For a slightly larger asymmetry, $\sigma_1=0.5$ and $\sigma_2=0.25$, we obtain stable equilibria (in Fig. \ref{figure_4}(b)). With this, the dynamical effects of asymmetries in the interactions between the layers of the network are explicitly demonstrated for this system.    

Hence, we can conclude that regular solutions are induced in the system as a result of mutual disturbances between the layers. But what is exactly the mechanism behind the newly found AiO and how is it related to synchronization? To address these questions, we first point out that in the absence of patch heterogeneities ($\gamma=0$) and layers asymmetries ($\sigma_1=\sigma_2$), the synchronization manifold ${\bf S}$ of the system, defined as ${\bf r}_{1\mu}={\bf r}_{2\mu}=\dots={\bf r}_{N\mu}={\bf r}_{1\mu'}={\bf r}_{2\mu'}=\dots={\bf r}_{N\mu'}$, follows a chaotic attractor ${\bf A}_{\bf S}$ embedded in the system's $3N$-dimensional state-space. However, the combinations of different dispersal constants in each layer generate an asymmetric inter-layer coupling term, with nonzero amplitude, that prevents the flow to visit the system's synchronization manifold ${\bf S}$. Hence, the system's trajectories cannot dwell on the chaotic attractor ${\bf A}_{\bf S}$ and, therefore, they visit alternative sets in the system's state-space that can eventually be nonchaotic.

\begin{figure}[!htp] 
\centering
\includegraphics[width=8.5cm,height=3.6cm]{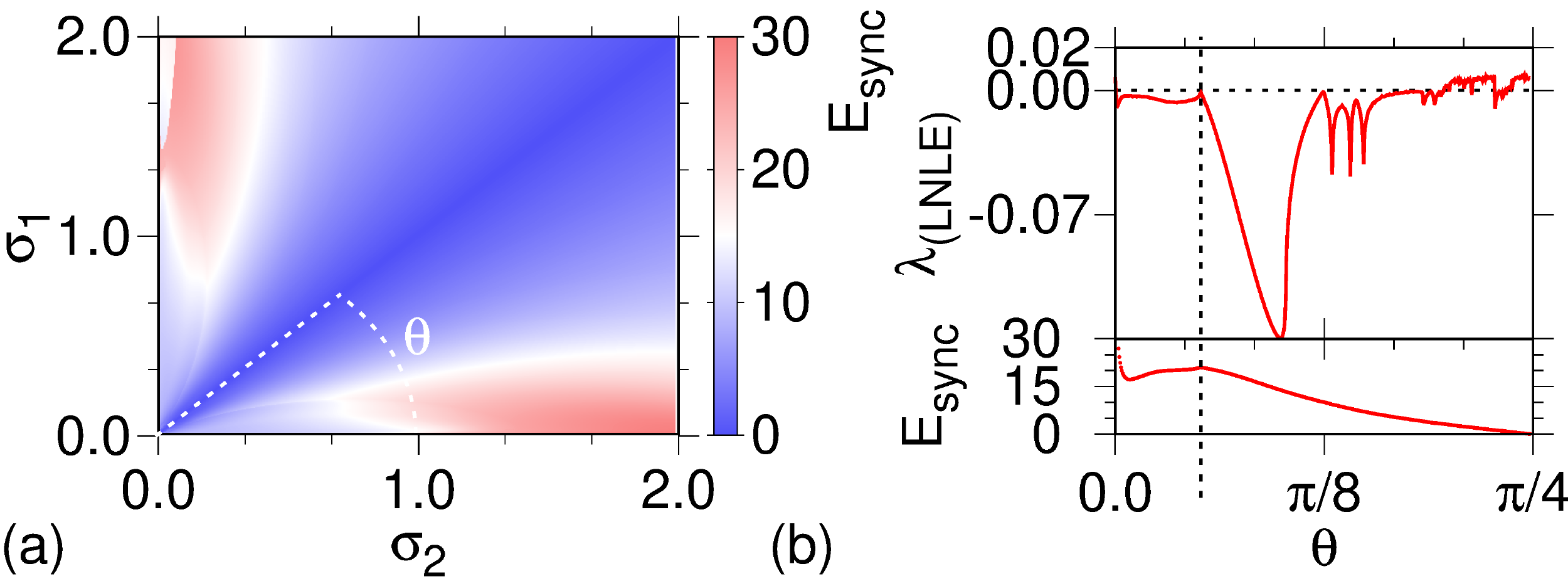}
\caption{(a) Synchronization error $E_{sync}$ in the ($\sigma_2$, $\sigma_1$) parameter plane. (b) The LNLE for the coupled system as a function of angle $\theta$ characterizing the level of topological asymmetry in the network, and the corresponding synchronization error $E_{sync}$.}
\label{figure_5}
\end{figure}

To visualize the relationship between AiO and the fact that the flow doesn’t remain on the synchronization manifold ${\bf S}$, we calculate the synchronization error among the patches $E_{sync}=\lim_{\tau\to\infty}[1/\tau\int_{0}^{\tau}\|\mathbf{r}_{(j+1)}-\mathbf{r}_{j} \|dt]$, where $j=i+(\mu-1)N_{\mu}$ with $i=1,\dots,N_{\mu}$ and $\mu=1,2$ (Fig. \ref{figure_5}(a)). Even for a slightly heterogeneous system ($\gamma=0.00015$), the system approaches synchronization ($E_{sync}\rightarrow 0$) in the symmetry axis of the diagram ($\sigma_1\rightarrow\sigma_2$). By comparing Fig. \ref{figure_5}(a) with Fig. \ref{figure_3}(a), we observe the correlation between chaotic solutions and the vanishing synchronization error ($E_{sync}\rightarrow 0$). This fact is elucidated by estimating LNLE and $E_{sync}$ as a function of the angle $\theta$ quantifying the asymmetry among the dispersal constants in the layers (Fig. \ref{figure_5}(b)). For $\theta=0$, layer $1$ is internally disconnected ($\sigma_1=0$). Consequently, chaotic oscillations occurring in the uncoupled patches cause $E_{sync}$ to be large. By varying $\theta$ in the interval $[0,\pi/4]$, for values slightly higher than zero, we observe that $E_{sync}$ drops significantly to a local minimum and subsequently increases to a local maximum (dashed line in Fig. \ref{figure_5}(b)) at which an inverse Hopf bifurcation provides a stable equilibrium. For further increasing $\theta$, the system approaches synchronous configurations ($E_{sync} \mapsto 0$) and the LNLE becomes positive. Consequently, regions corresponding to regular solutions become smaller and sparse giving room to intervals of parameters leading to chaotic behavior as the angle $\theta$ approaches $\pi/4$. In fact, for $\theta > \pi/5$, chaotic solutions are already predominant with the occurrence of small periodic windows. 

Furthermore, for homogeneous patches ($\gamma=0$), the impact of dispersal asymmetries on synchronization can also be analytically demonstrated by expressing the equations of each layer as:
\begin{eqnarray}
 \label{sys2}
 \nonumber
 {\bf \dot{\;L}}_1 &=& {\bf F}({\bf L}_{1}) + \sigma_1({\bf G}\otimes{\bf H}){\bf L}_{1} + ({\bf \mathcal{A}}\otimes{\bf H})(\sigma_2{\bf L}_2-\sigma_1{\bf L}_1), \\
 {\bf \dot{\;L}}_2 &=& {\bf F}({\bf L}_{2}) + \sigma_2({\bf G}\otimes{\bf H}){\bf L}_{2} - ({\bf \mathcal{A}}\otimes{\bf H})(\sigma_2{\bf L}_2-\sigma_1{\bf L}_1),
 \nonumber
\end{eqnarray}
where $\mathbf{L}_1=[{\bf r}_{11},{\bf r}_{21},\dots,{\bf r}_{N_11}]$ and $\mathbf{L}_2=[{\bf r}_{12},{\bf r}_{22},\dots,{\bf r}_{N_22}]$ are the state vectors of the layers. The function ${\bf F}({\bf L}_{\mu})$ describes the dynamics of individual patches. The $N_{\mu} \times N_{\mu}$ matrix ${\bf G}$ specifies the intra-layer connectivity, while the $N_{\mu} \times N_{\mu'}$ matrix $\mathcal{A}$ determines the inter-layer connections. The $3 \times 3$ matrix ${\bf H}$ defines variables of ${\bf F}({\bf L}_{\mu})$ engaged in the intra- and inter-layer interactions (see Appendix A for these matrices). With this, the stability of dynamics on $\mathbf{S}$ would imply $\mathbf{L}_1=\mathbf{L}_2$, therefore, $\delta\mathbf{\dot{\;L}} = \mathbf{\dot{\;L}}_1-\mathbf{\dot{\;L}}_2=0$. Hence,
\begin{eqnarray}
\label{condition}
 \delta\mathbf{\dot{\;L}}=(\sigma_1-\sigma_2)[(\mathbf{G}-2\mathcal{A})\otimes\mathbf{H}]\mathbf{L}_1=0.
\end{eqnarray}
Since the intra- and inter-layer connectivities are different, for nonzero biomass in the plants and animal species, the condition in Eq. (\ref{condition}) is satisfied only if $\sigma_1=\sigma_2$. Therefore, the synchronization manifold does not support the flow for $\sigma_1 \neq \sigma_2$, confirming the correlation between AiO and the absence of dynamics on $\mathbf{S}$. 

The AiO phenomenon relies on the absence of dynamics on the synchronization manifold. Hence, we investigate its sensitivity to the system characteristics that enhance or suppress synchronization. More precisely, we vary the patch's heterogeneities defined by $\gamma$, and the number of inter-layer links $l \in [0,N_{\mu}]$. First, we introduce a measure to estimate the fraction of dispersal constant combinations that leads to non-chaotic solutions. Formally, we define a finite region of the stability diagram in the ($\sigma_2$, $\sigma_1$) parameter plane as $\mathcal{E}_{\gamma l} = \{ (\sigma_2,\sigma_1) \in \mathbb{R}^2 \mid \sigma_2 \in [0,5], \sigma_1 \in [0,5] \}$. The combination of dispersal constants resulting in regularity is defined as the subset of $\mathcal{E}_{\gamma l}$ at which $\mathbf{\lambda}_{LNLE}<0$, and denoted by $\mathcal{S}_{\gamma l}$. With this, we estimate the relative portion of the stability diagram leading to non-chaotic solutions as the intersection of $\mathcal{S}_{\gamma l}$ and $\mathcal{E}_{\gamma l}$:
\begin{equation}
 RI_{\gamma l}=\frac{Vol(\mathcal{S}_{\gamma l} \cap \mathcal{E}_{\gamma l})}{Vol(\mathcal{E}_{\gamma l})}.
\end{equation}
We call this measure regularity-index (RI). The subscripts $\gamma$ and $l$ stand for the patch heterogeneity and the number of inter-layer links, respectively.

\begin{figure}[!htp] 
\centering
\includegraphics[width=8.5cm,height=3.6cm]{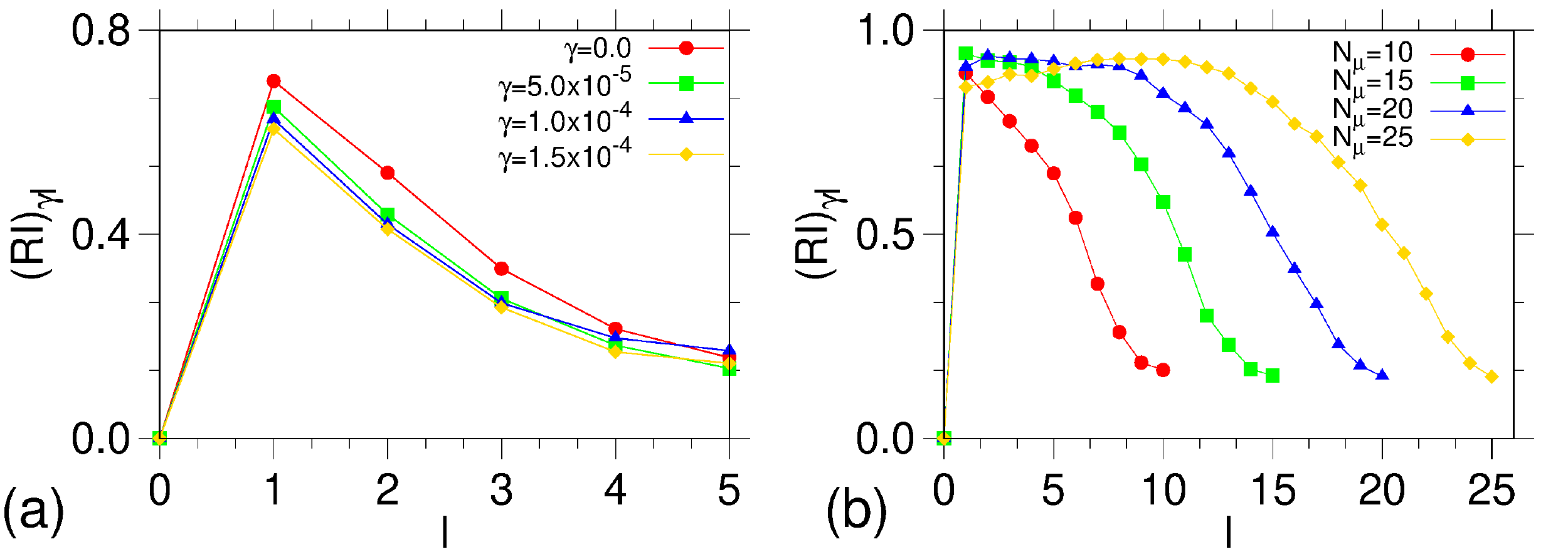}
\caption{The regularity-index $RI_{\gamma l}$ as a function of the number of inter-layer links $l$. (a) For different levels of  heterogeneity in the patches $\gamma$ ($N_{\mu}=5$). (b) For different layer sizes $N_{\mu}$ ($\gamma=0$).}
\label{figure_6}
\end{figure}

Interestingly, for $N_{\mu}=5$, AiO is optimal if only one inter-layer link is present (Fig. \ref{figure_6}(a)). This agrees with the fact that a high number of inter-layer links would facilitate synchronization and, therefore, stimulate chaotic solutions. Comparing the results for different levels of heterogeneity in the patches, specified by $\gamma$ in Fig. \ref{figure_6}(a), we observe that AiO is more pronounced for the homogeneous patches ($\gamma=0$). As the onset of AiO is related to perturbations only in the patches involved in the inter-layer interaction, such disturbance should percolate the entire layer in a finite time, which is hindered by patch heterogeneities. This effect becomes weaker for larger values of $l$ since the inter-layer disturbances are less localized. Nevertheless, the AiO is manifested for all levels of patch heterogeneities considered here. Besides, for cases in which the layers exhibit complex topologies, the topological relevance of individual nodes \cite{Tlaie2019} may also be considered. In Fig. \ref{figure_6}(b), we remark the occurrence of AiO for larger networks.

\section{Conclusions}

In conclusion, we report the onset of stable solutions with low amplitude and low temporal variability due to topological asymmetry of multilayer networks. We call this phenomenon asymmetry-induced order (AiO). The AiO is related to the absence of dynamics on the system's synchronization manifold, an invariant set at which chaotic solutions with high variability dwell. A comprehensive analysis of the system stability diagrams reveals that a minimal number of inter-layer links and patches homogeneity are optimal conditions for AiO. Here, AiO has been numerically exemplified in a model for the dispersal of species throughout layered landscapes. However, we also observe AiO in numerical simulations of networks composed by the paradigmatic Lorenz and R\"ossler systems (Appendix B). This suggests that this phenomenon is a general characteristic of diffusive processes in networks. Therefore, AiO may also play a role in the diffusion of chemical reactions, propagation of contagious diseases, the transmission of electrical signals in the brain, and the heart.

\section{Acknowledgments}
\begin{acknowledgments}
This work was supported by the Deutsche Forschungsgemeinschaft (DFG) (Projects: 163436311-SFB-910, INST-184/157-1 FUGG, FE359/20-1 FOR2716) and the MWK of the State Lower Saxony for the High-Performance Computer CARL.
\end{acknowledgments}

\appendix

\section{Appendix A: Matrices defined in the main text}

The Laplacian matrix ${\bf G}$ describing the intra-layer connections is given by:

\begin{equation}
 {\bf G}=
 \begin{pmatrix}
-2 & 1 & 0 & 0 & 1 \\
1 & -2 & 1 & 0 & 0 \\
0 & 1 & -2 & 1 & 0 \\
0 & 0 & 1 & -2 & 1 \\
1 & 0 & 0 & 1 & -2
\end{pmatrix}
\label{intra}
\end{equation}

The $N_{\mu} \times N_{\mu`}$ matrix $\mathcal{A}$ specifying the inter-layer connections is given by: 
\begin{equation}
\mathcal{A}=
 \begin{pmatrix}
0 & 0 & 0 & 0 & 0 \\
0 & 1 & 0 & 0 & 0 \\
0 & 0 & 1 & 0 & 0 \\
0 & 0 & 0 & 0 & 0 \\
0 & 0 & 0 & 0 & 0
\end{pmatrix}
\end{equation}

The $3 \times 3$ matrix ${\bf H}$ also defined in the main text specifies the components of the state ${\bf r}$ engaged in the intra- and inter-layer connections. Explicitly, this matrix reads as:
\begin{equation}
\mathbf{H}=
 \begin{pmatrix}
0 & 0 & 0 \\
0 & 1 & 0 \\
0 & 0 & 1
\end{pmatrix}
\end{equation}

\section{Appendix B: Asymmetry-induced order in double-layer networks of paradigmatic models}

The dynamical consequences of breaking the coupling symmetry in a double-layer network are also visible for other nonlinear systems as network units. In the following subsections, we show the results for two paradigmatic systems. 

\subsection{R\"ossler System}

For the R\"ossler system, the initial value problem $\dot{{\bf r}}={\bf f}({\bf r})$ with ${\bf r}(0)={\bf r}_0$ can be explicitly written as:
\begin{eqnarray}
\label{Rossler}
\nonumber
\dot{x} &=& - y - z,\\ [5pt]
\dot{y} &=& x + ay,\\ [5pt]
\dot{z} &=& b + z(x - c),
\nonumber
\end{eqnarray}
where ${\bf r}(t)=(x(t), y(t), z(t))$. The control parameters are fixed at $a=0.42$, $b=2.0$, and $c=3.94$. The $N_{\mu} \times N_{\mu}$ matrix ${\bf G}$ specifies the intra-layer connectivity (first neighbors), see matrix (\ref{intra}). The $N_{\mu} \times N_{\mu`}$ matrix $\mathcal{A}$ specifies the inter-layer connections as: 
\begin{equation}
\mathcal{A}=
 \begin{pmatrix}
0 & 0 & 0 & 0 & 0 \\
0 & 1 & 0 & 0 & 0 \\
0 & 0 & 0 & 0 & 0 \\
0 & 0 & 0 & 0 & 0 \\
0 & 0 & 0 & 0 & 0
\end{pmatrix}
\end{equation}

The $3 \times 3$ matrix ${\bf H}$ specifies the components of the state ${\bf r}$ engaged in the intra and inter-layer connections:
\begin{equation}
\mathbf{H}=
 \begin{pmatrix}
0 & 0 & 0 \\
0 & 1 & 0 \\
0 & 0 & 1
\end{pmatrix}
\end{equation}

\begin{figure}[!htp] 
\centering
\includegraphics[width=7cm,height=5.5cm]{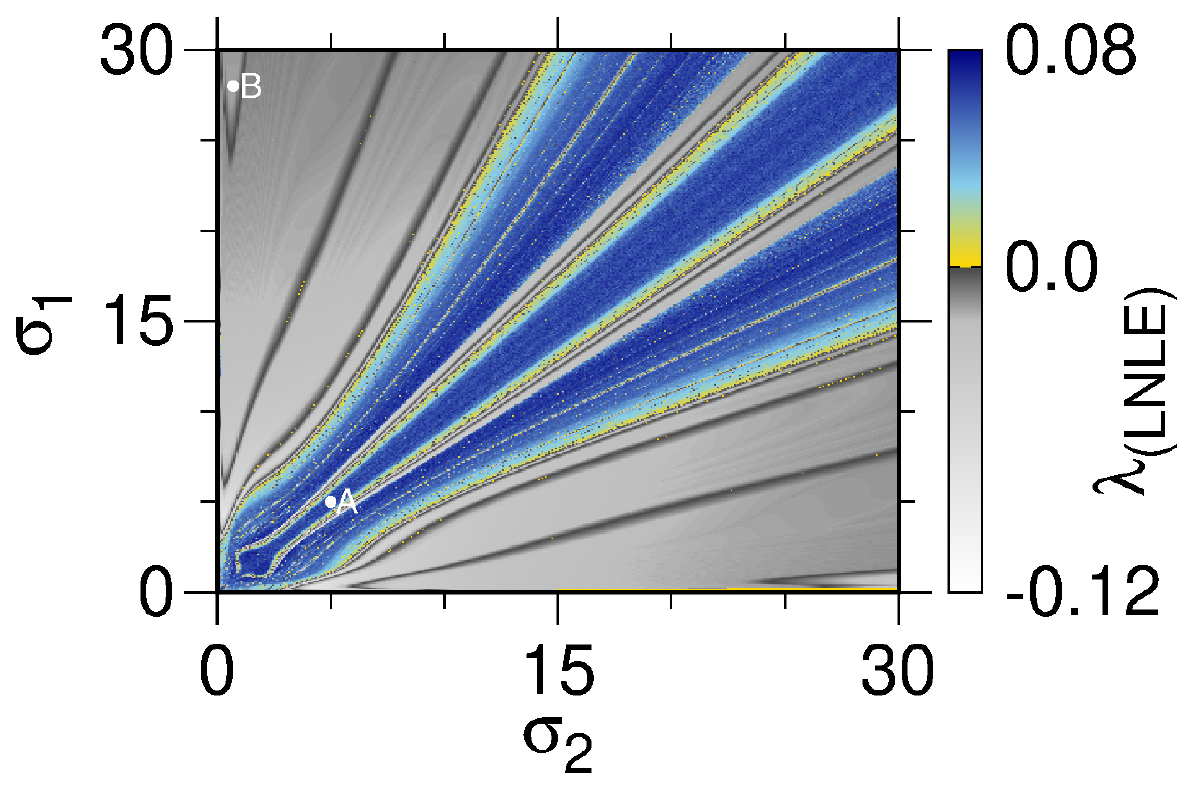}
\caption{Two-dimensional ($\sigma_2 \times \sigma_1$) stability diagram. The color-code represents the LNLE ($\lambda_{(LNLE)}$). Positive values of $\lambda_{(LNLE)}$ indicate chaotic solutions, while negative values stand for regular solutions.}
\label{figure_appendix1}
\end{figure}

In Fig. \ref{figure_appendix1}, the regularity of solutions in this R\"ossler network is assessed using the largest nonzero Lyapunov exponents (LNLE) \cite{Wolf1985} for a large combinations of $\sigma_1$ and $\sigma_2$. 
The occurrence of regular solutions ($\lambda_{LNLE}<0$) off the diagonal axis of the diagram in Fig. \ref{figure_appendix1} indicate the onset of asymmetry-induced order for this double-layer network.

\begin{figure*}[!htp] 
\centering
\includegraphics[width=12cm,height=6cm]{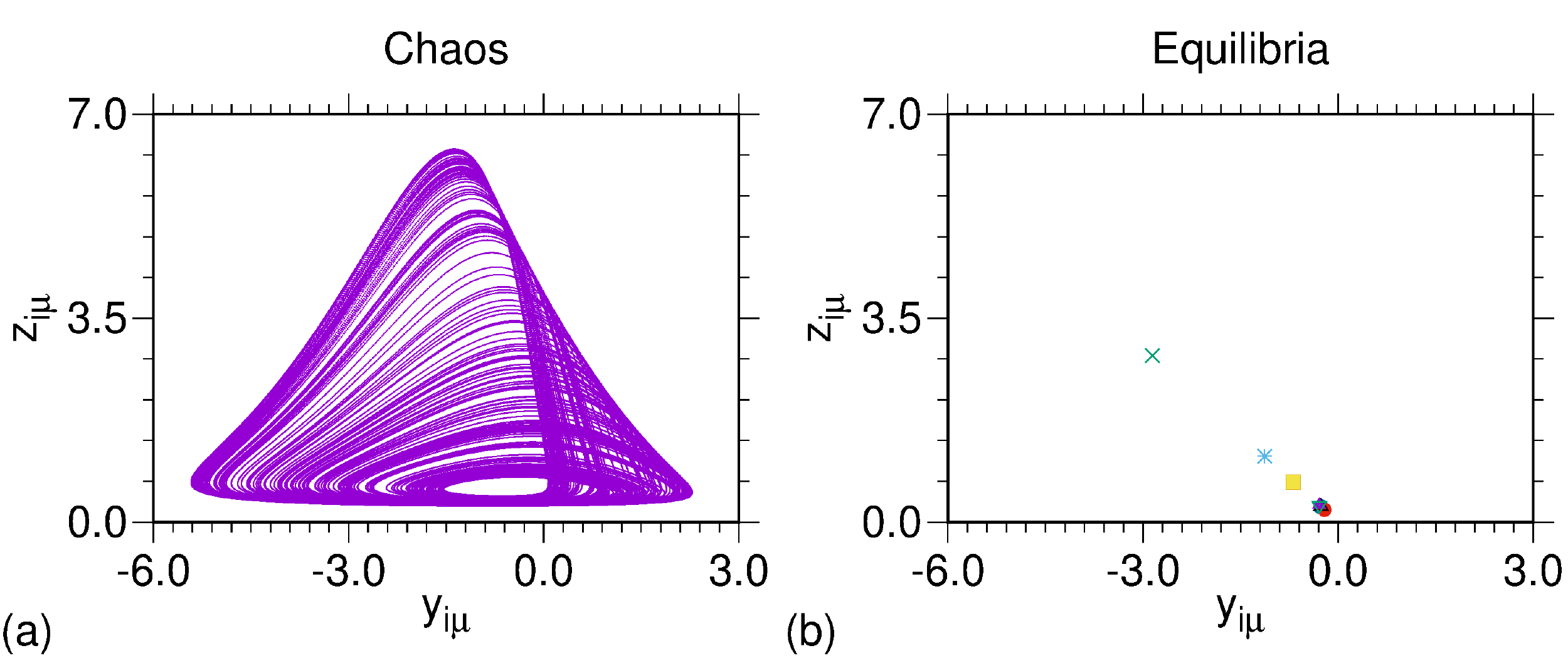}
\caption{(a) State-space projections of chaotic solutions occurring for symmetric configurations of the layers, i.e., $\sigma_1=\sigma_2=5.0$ (point $A$ in Fig. \ref{figure_appendix1}). (b) State-space of equilibria solutions occurring for $\sigma_1=28$ and $\sigma_2=0.7$. (point $B$ in Fig. \ref{figure_appendix1}). }
\label{figure_appendix2}
\end{figure*}

In Fig. \ref{figure_appendix2}, we show a state-space projection of solutions obtained for constants ($\sigma_1$,$\sigma_2$) indicated by the points $A$ and $B$ in Fig. \ref{figure_appendix1}. As reported in the main text for a spatially extended food-web, symmetric combinations of $\sigma_1$ and $\sigma_2$ yield chaotic solutions (Fig. \ref{figure_appendix2}(a)), while asymmetric combinations result in stable equilibria (Fig. \ref{figure_appendix2}(b)). These observations confirm the occurrence of AiO for a double-layer network of R\"ossler units.

\subsection{Lorenz System}

For the Lorenz system, the state vector ${\bf r}(t)$ is prescribed by:
\begin{eqnarray}
\label{Lorenz}
\nonumber
\dot{x} &=& \sigma(y-x),\\ [5pt]
\dot{y} &=& x(\rho-z)-y,\\ [5pt]
\dot{z} &=& xy-\beta z,
\nonumber
\end{eqnarray}
where ${\bf r}(t)=(x(t), y(t), z(t))$. The control parameters are fixed at the typical values for chaos in the Lorenz system, i.e., $\sigma=10$, $\rho=28$, and $\beta=8/3$. The $N_{\mu} \times N_{\mu}$ matrix ${\bf G}$ specifies the intra-layer connectivity (first neighbors), see matrix (\ref{intra}). The $N_{\mu} \times N_{\mu`}$ matrix $\mathcal{A}$ specifies the inter-layer connections as: 
\begin{equation}
\mathcal{A}=
 \begin{pmatrix}
1 & 0 & 0 & 0 & 0 \\
0 & 1 & 0 & 0 & 0 \\
0 & 0 & 1 & 0 & 0 \\
0 & 0 & 0 & 0 & 0 \\
0 & 0 & 0 & 0 & 0
\end{pmatrix}
\end{equation}

The $3 \times 3$ matrix ${\bf H}$ specifies the components of the state ${\bf r}$ engaged in the intra and inter-layer connections:
\begin{equation}
\mathbf{H}=
 \begin{pmatrix}
1 & 0 & 0 \\
0 & 1 & 0 \\
0 & 0 & 1
\end{pmatrix}
\end{equation}

\begin{figure}[!htp] 
\centering
\includegraphics[width=7cm,height=5.5cm]{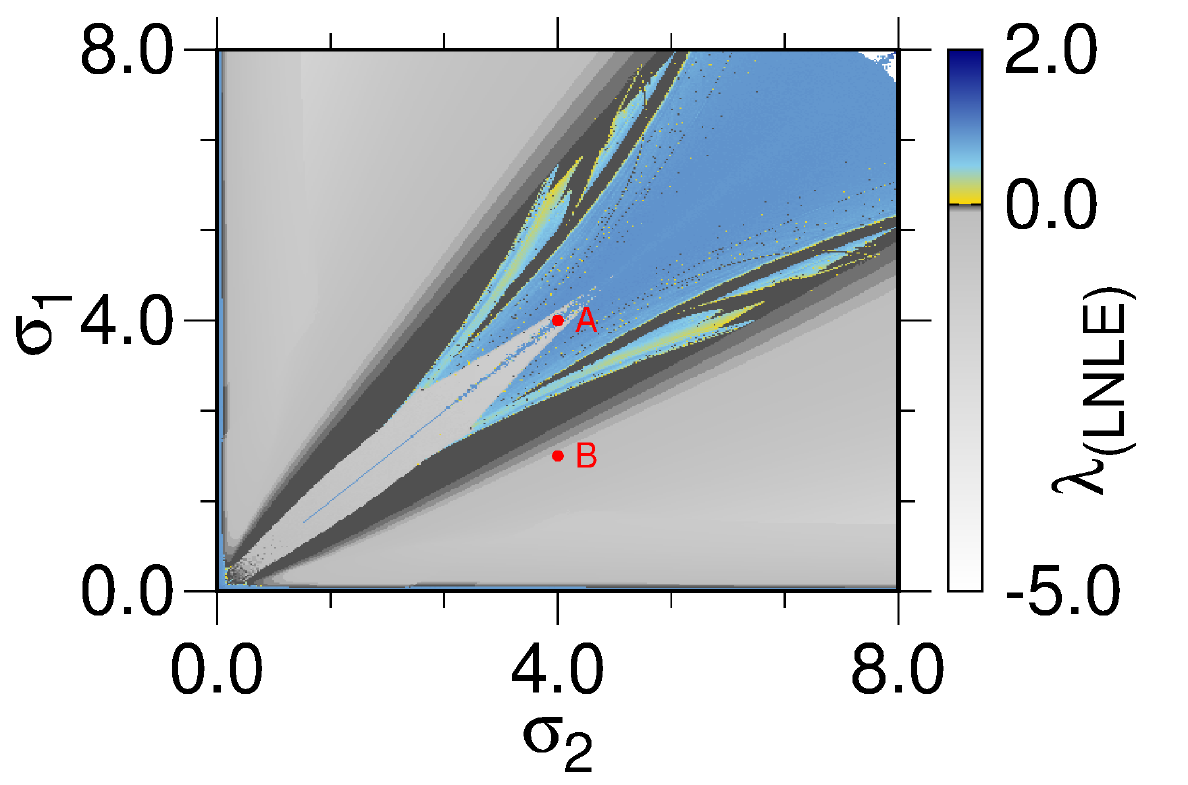}
\caption{Two-dimensional ($\sigma_2 \times \sigma_1$) stability diagram. The color-code represents the LNLE ($\lambda_{(LNLE)}$). Positive values of $\lambda_{(LNLE)}$ indicate chaotic solutions, while negative values stand for regular solutions.}
\label{figure_appendix3}
\end{figure}

In Fig. \ref{figure_appendix3}, the regularity of solutions in this Lorenz network is also assessed using the (LNLE \cite{Wolf1985} for combinations of $\sigma_1$ and $\sigma_2$.

In this diagram, regular solutions ($\lambda_{LNLE}<0$) are visualized for asymmetric combinations of the coupling strengths, therefore, the occurence of asymmetry-induced order is confirmed. In particular for the Lorenz system, for a small interval of symmetric combinations of $\sigma_1$ and $\sigma_2$, the synchronization manilfold is unstable. As a consequence, regular solutions with ($\lambda_{LNLE}<0$) also appears for $\sigma_1=\sigma_2$ in the interval $\sim$ [$0.1$,$1.0$].  

\begin{figure*}[!htp] 
\centering
\includegraphics[width=12cm,height=6cm]{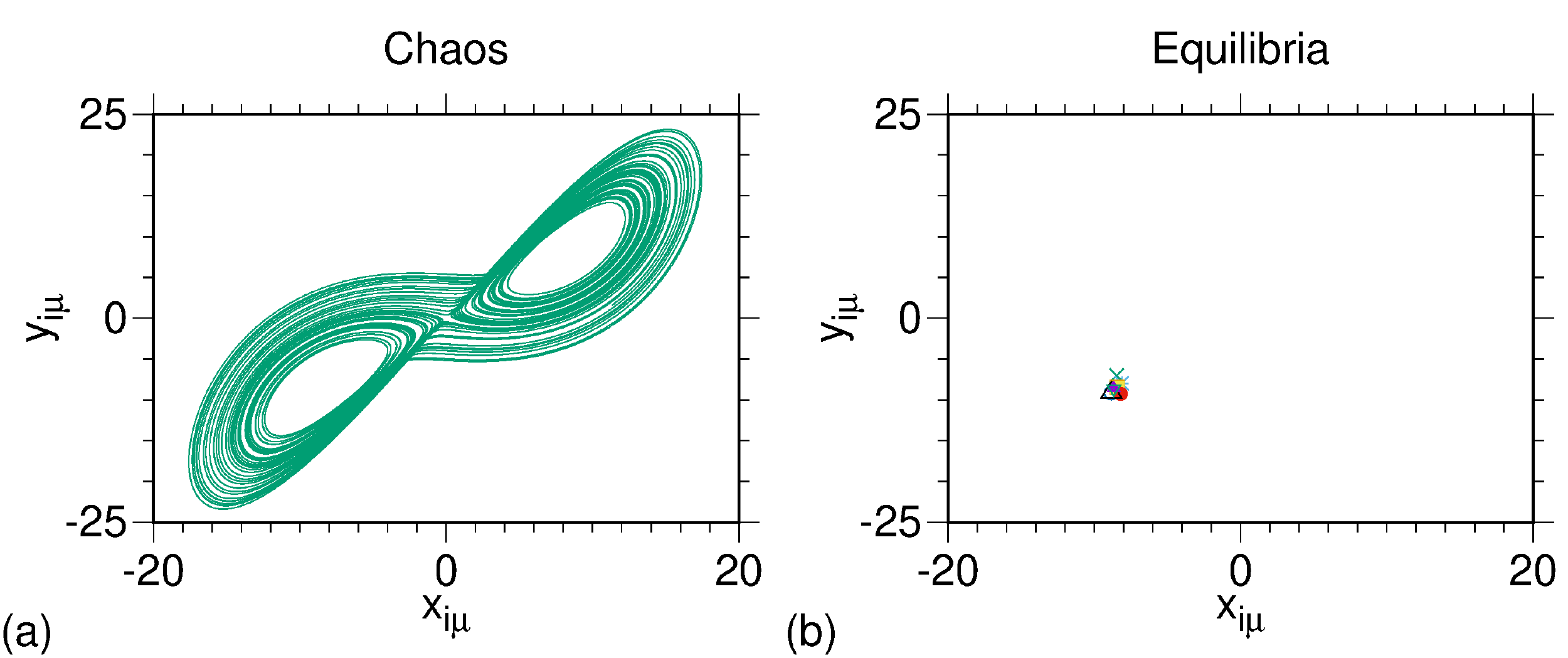}
\caption{(a) State-space projections of chaotic solutions occurring for symmetric configurations of the layers, i.e., $\sigma_1=\sigma_2=4.0$ (point $A$ in Fig. \ref{figure_appendix3}). (b) State-space of equilibria solutions occurring for $\sigma_1=4.0$ and $\sigma_2=2.0$. (point $B$ in Fig. \ref{figure_appendix3}). }
\label{figure_appendix4}
\end{figure*}

In Fig. \ref{figure_appendix4}, we show a state-space projection of solutions obtained for constants ($\sigma_1$,$\sigma_2$) indicated by the points $A$ and $B$ in Fig. \ref{figure_appendix3}. We observe for this case that most symmetric combinations of $\sigma_1$ and $\sigma_2$ yield chaotic solutions (Fig. \ref{figure_appendix4}(a)), while asymmetric combinations result in stable equilibria (Fig. \ref{figure_appendix4}(b)). These observations confirm the occurrence of Asymmetry-induced order (AiO) for a double-layer network of Lorenz units.   

%


\end{document}